\documentclass[12pt,a4paper]{article} %% final layout
\usepackage{hyperref}
\usepackage{ol2}
\usepackage{amsmath}

\begin{document}

%\twocolumn[ %% activate for two-column option

%\title{Extreme subwavelength nanocavity modes}
%\title{Extreme subwavelength nanocavity modes in {\it U-} shaped metal-insulator-metal surface plasmon polariton resonators}
\title{Plasmonic modes of extreme subwavelength nanocavities}

\author{J\"org Petschulat$^{1,*}$, Christian Helgert$^1$, Michael Steinert$^1$, Norbert Bergner$^1$, Carsten Rockstuhl$^{2}$, Falk Lederer$^{2}$, Thomas Pertsch$^{1}$, Andreas T\"unnermann$^{1}$, and Ernst-Bernhard Kley$^1$}
\address{
$^1$Institute of Applied Physics, Friedrich-Schiller-Universit\"at Jena, Max-Wien-Platz 1, 07743, Jena, Germany\\
$^2$Institute for Condensed Matter Theory and Solid State Optics, Friedrich-Schiller-Universit\"at Jena, Max-Wien-Platz 1, 07743, Jena, Germany\\$^*$Corresponding author: joerg.petschulat@uni-jena.de}

\begin{abstract} We study the physics of a new type of subwavelength nanocavities. They are based on {\it U}-shaped metal-insulator-metal waveguides supporting the excitation of surface plasmon polaritons. The waveguides are  simultaneously excited from both sides of the {\it U} by incident plane waves. Due to their finite length discrete modes emerge within the nanocavity. We show that the excitation symmetry with respect to the cavity ends permits the observation of even and odd modes. Our investigations include near and far field simulations and predict a strong spectral far field response of the comparable small nanoresonators. The strong near field enhancement observed in the cavity at resonance might be suitable to increase the efficiency of nonlinear optical effects, quantum analogies and might facilitate the development of active optical elements, such as active plasmonic elements.
\end{abstract}
 %]

\noindent The exploration of localized and guided electromagnetic waves at the interface between a metal and a dielectric, in form of localized (LSPPs) or propagating surface plasmon polaritons (PSPPs), was stimulated by various novel applications that came in reach \cite{Oulten2009, Stockman2008,Bozhevolnyi2006,Huang2008}. The essential ingredient in these applications is usually a plasmonic element which exhibits specific spectral resonances or which allows to confine the electromagnetic field in well-defined spatial domains. 
%A great share of research was devoted to propose new plasmonic elements \cite{Oulten2009, Stockman2008,Bozhevolnyi2006,Huang2008}.

Here we introduce a novel element which belongs to the class of plasmonic nanocavities. The building blocks of PSPP waveguides are usually metallic wire elements infinitely extended in the longitudinal dimension. By choosing a suitable geometry in the transverse direction, various types of waveguides were proposed. Examples are waveguides for gap \cite{Miyazaki2006}, channel \cite{Bozhevolnyi2006,Orenstein2007} and wedge \cite{Vidal2008} plasmon polaritons. 
Nanocavities are typically obtained by cutting the waveguide to a finite, very short length \cite{Nordlander2009,Hu2009}.  In particular cases the resonant elements are represented by single particles, inter-particle gaps \cite{Nordlander2009} or metal voids where the PSPP is guided along predefined circuits \cite{Bozhevolnyi2006}. %der satz nicht so ganz klar; irgendwie mŸsste man doch sagen, wie nun die cavities aus den wellenleitern geformt werden; bei single particle kann man sich das ja noch vorszellem aber what the heck sind inner particle gaps

%\begin{figure}[htb]
%\centerline{\includegraphics[width=5.5cm]{fig1}}
%\caption{(Color online) The different steps of the proposed fabrication process. The process starts with\textbf{(a)}, the fabrication of an initial $Al$ wire grating. It is followed by the oxidation process \textbf{(b)}. In a last step \textbf{(c)}, the entire structure is metallized to obtain the {\it U-} shaped nanocavities.}
%\label{fig1}
%\end{figure}

\begin{figure}[p]
\centerline{\includegraphics[width=15cm]{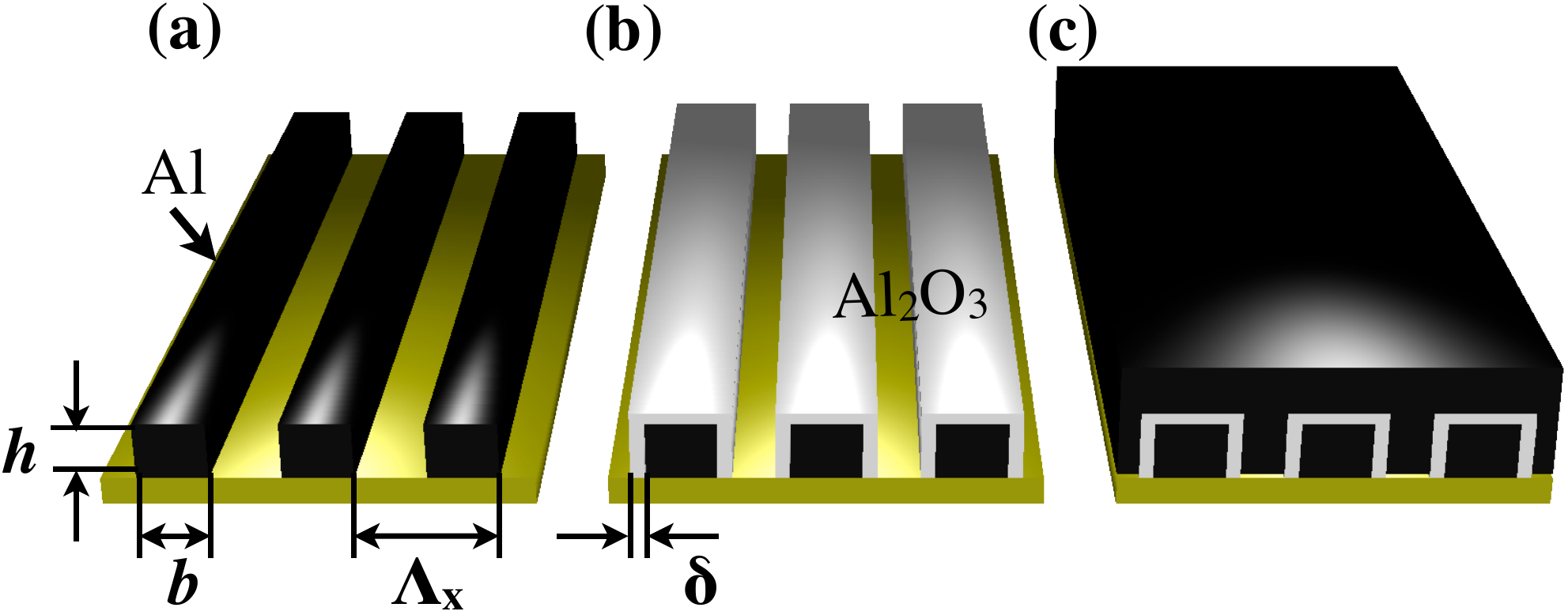}}
\caption{(Color online) The different steps of the proposed fabrication process. The process starts with\textbf{(a)}, the fabrication of an initial $Al$ wire grating. It is followed by the oxidation process \textbf{(b)}. In a last step \textbf{(c)}, the entire structure is metallized to obtain the {\it U-} shaped nanocavities.}
\label{fig1}
\end{figure}

The nanocavity we investigate here is formed in a planar, one-dimensionally confined metal-insulator-metal (MIM) waveguide. The advantage of the MIM setup is a high confinement of the plasmon energy in the dielectric \cite{Polman2008} that even allows to bend the waveguide \cite{Leen2007}, without excessive radiation losses as it would be observed for insulator-metal-insulator (IMI) waveguides \cite{Zon2007,Smith2007}. We fabricated such structures employing a self-limiting process that relies on the intrinsic oxidation of bulk aluminum. This yields a solid $1-3~\text{nm}$ (depending on the environmental parameters) thin aluminum-oxide barrier layer ($Al_2O_3$), preventing the $Al$ underneath from any further oxidation. The formation of such a layer is well documented and its optical properties have been investigated \cite{Langhammer2008}. The fabrication steps are summarized in Fig.\ref{fig1}. It starts with the oxidization of rectangular-shaped stripe gratings made of $Al$. The native oxidation process, that might be additionally supported by an oxygen plasma, results in an ultrathin $Al_2O_3$ covering of the $Al$ wires. In a final step the oxidized wires are embedded in $Al$ to form a symmetric MIM waveguide. This fabrication process combines the deterministic electron beam lithography with the self-limiting natural oxidation of $Al$.

\begin{figure}[p]
\centerline{\includegraphics[width=15cm]{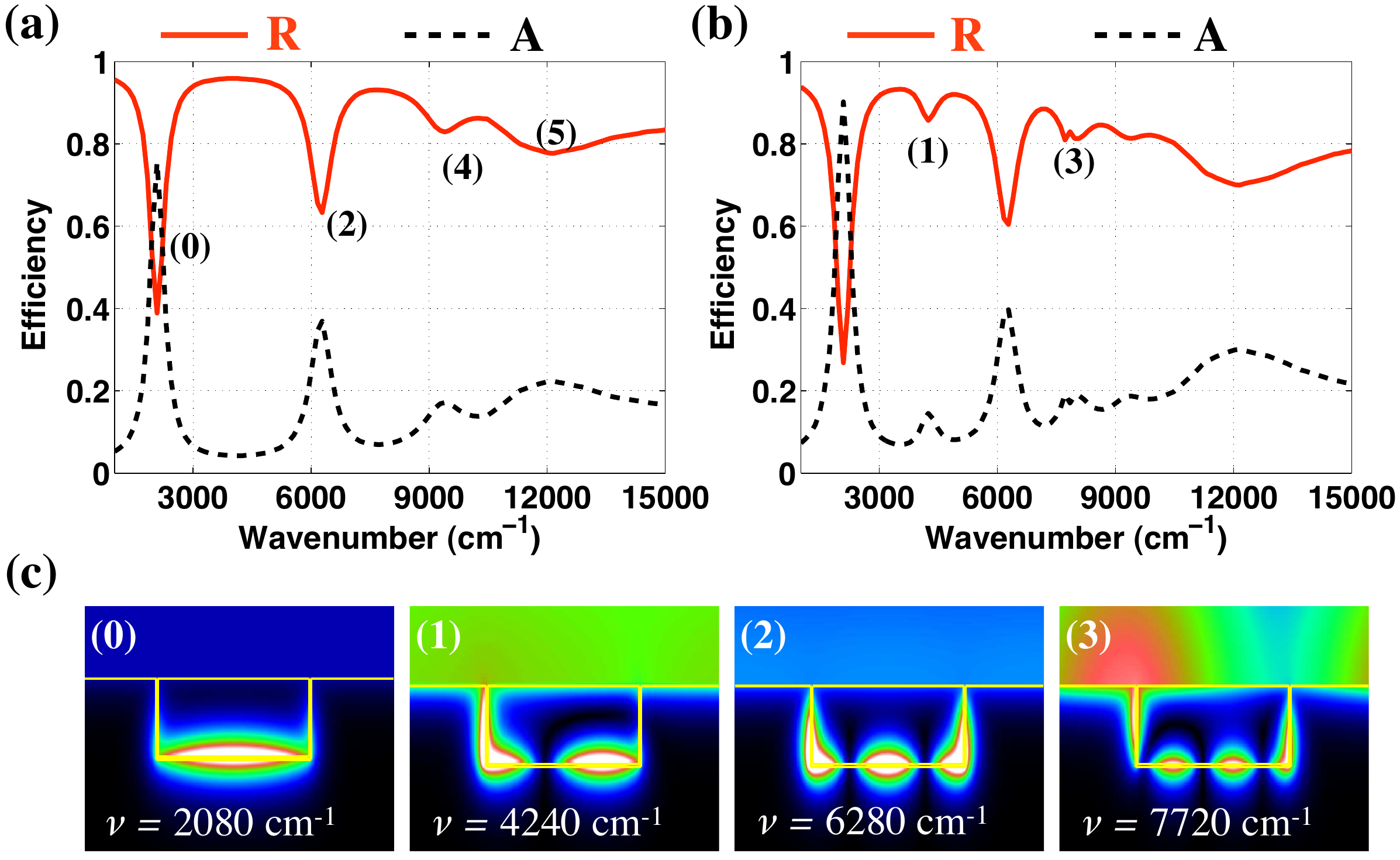}}
\caption{(Color online) Calculated far field spectra (reflection and absorption, transmission is zero) for \textbf{(a)} normal and \textbf{(b)} oblique ($45^\circ$) incidence. \textbf{(c)} Simulation of the modulus of the magnetic field for spectrally observed first four modes (0-3).}
\label{fig2}
\end{figure}

In order to investigate the possible modes propagating inside such cavities we performed numerical simulations of the structure above, taking realistic parameters for the geometry into account, amenable for the fabrication. For the resonator width $b$ we choose $250~\text{nm}$, for the height $h=60~nm$ yielding a total cavity length of $370~\text{nm}$. We set the period within the periodic arrangement of the structures to $\Lambda_x=500~\text{nm}$ and the oxide layer thickness to $\delta=2.28~\text{nm}$. This value has been determined via X-ray diffraction measurements on a flat $Al$ film covered by the native $Al_2O_3$ layer. To obtain the far field spectra as well as the near field response, shown in Fig.\ref{fig2} the Fourier Modal Method \cite{Lie1997} was used. In our simulation we took the dispersion of $Al$ fully into account. The substrate material was assumed to be a dielectric with $n_\text{sub}=1.5$. In order to excite MIM modes we choose TM (p-) polarized incident light.

%\begin{figure}[htb]
%\centerline{\includegraphics[width=8.0cm]{fig2}}
%\caption{(Color online) Calculated far field spectra (reflection and absorption, transmission is zero) for \textbf{(a)} normal and \textbf{(b)} oblique ($45^\circ$) incidence. \textbf{(c)} Simulation of the modulus of the magnetic field for spectrally observed first four modes (0-3).}
%\label{fig2}
%\end{figure}

Results for the reflected and absorbed intensity upon normal incidence are shown in Fig.~\ref{fig2}a. Four different resonances may be unambiguously resolved. Resonances indicated by (0,2,4) represent the fundamental, the second and the fourth order nanocavity mode, respectively. Resonances were labeled according to the symmetry of the magnetic field with respect to the center of the nanocavity unit cell. This can be deduced from the simulated near fields, which are shown in Fig.~\ref{fig2}c. By contrast, resonance (5) is related to the interband transition of bulk $Al$ \cite{Langhammer2008} and is not induced by the nanocavity. This can be revealed from more systematic simulations where the geometrical parameters of the nanocavity are modified. We mention that the ratio of gap width to resonance wavelength ($\delta / \lambda$) is in the order of $1/1000$. By increasing the angle of incidence and illuminating the structure obliquely, in addition to the even modes (0,2,4) the missing odd modes (1,3), which were not excitable before due to symmetry constraints, can be observed. The spectral response for an illumination angle of $45^\circ$ is shown in Fig.~\ref{fig2}b, near fields are show in Fig.~\ref{fig2}c together with the even near field distributions.

%\begin{figure}[htb]
%\centerline{\includegraphics[width=8.0cm]{fig3}}
%\caption{(Color online) \textbf{(a)} The calculated far-field spectra for two {\it U} shaped cavity with different shape, but with the same total cavity length (lines and circles). The initial structure as presented in Fig. \ref{fig2} is represented by the solid and dashed lines. \textbf{(b)} The analytical resonance dependence (dashed line) according to Eq. (\ref{eq1}) together with the numerically estimated resonance positions (triangles and squares).}
%\label{fig3}
%\end{figure}

\begin{figure}[p]
\centerline{\includegraphics[width=15cm]{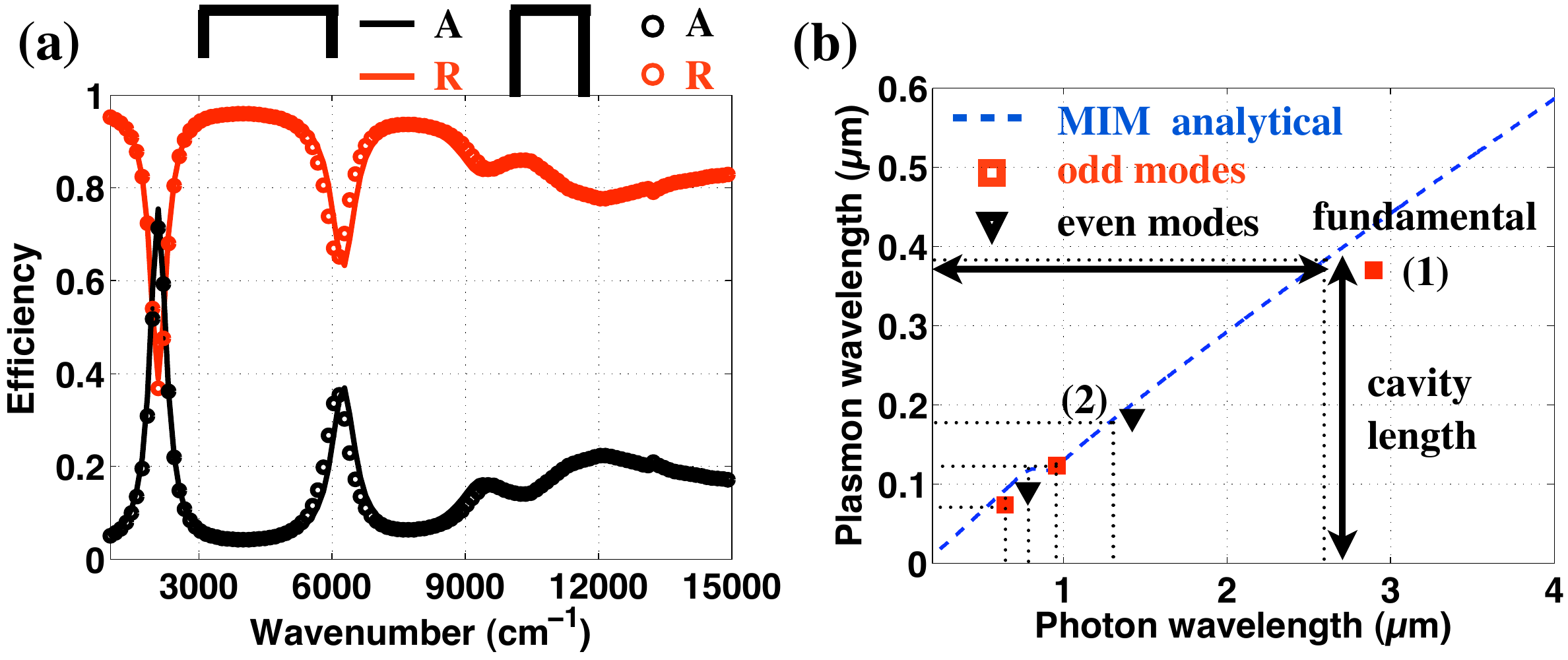}}
\caption{(Color online) \textbf{(a)} The calculated far-field spectra for two {\it U} shaped cavity with different shape, but with the same total cavity length (lines and circles). The initial structure as presented in Fig. \ref{fig2} is represented by the solid and dashed lines. \textbf{(b)} The analytical resonance dependence (dashed line) according to Eq. (\ref{eq1}) together with the numerically estimated resonance positions (triangles and squares).}
\label{fig3}
\end{figure}

In order to verify that the entire waveguide circumference acts as a nanocavity, we performed additional simulations for different nanocavity heights and widths, but with the restriction that the entire cavity length ($2h+b$) remains constant. For a selected set of parameters ($h=110~\text{nm};~b=150~\text{nm}$) the spectrum is compared to the spectrum obtained with the original parameters. As can be seen in Fig. \ref{fig3}a, both configuration yield an identical response despite a substantial difference in the resonator shape. This suggests that the sharp corners in each nanocavity do not influence the resonance positions which are solely determined by the cavity length and the particular oxide layer thickness and material.\\

In order to categorize the guided modes as MIM surface plasmon polariton modes, we solve the dispersion relation for the MIM configuration \cite{Orenstein2007}
\begin{eqnarray}
\text{tanh}\left(\frac{k_d(\omega)\delta}{2}\right)&=&\frac{\varepsilon_d(\omega)
k_m(\omega)}{\varepsilon_m(\omega)k_d(\omega)},\nonumber \\
k_{m,d}(\omega)&=&\sqrt{k_\text{gap}^2(\omega)-\varepsilon_{m,d}(\omega)
\frac{\omega_2}{c_2}}.
\label{eq1}
\end{eqnarray}
Eq. (\ref{eq1}) allows to calculate the effective plasmon wavelength depending on the vacuum wavelength of the illumination. Here $\varepsilon_{m,d}$ denotes the dielectric function of the metal ($Al$) and gap material ($Al_2O_3;~\varepsilon=2.756$ in the visible spectra), $\delta$ is the gap distance, $\omega$ the frequency and $c$ the vacuum speed of light. Considering the nanocavity, one would expect that the finite cavity length divided by integer numbers according to the mode order represents the condition for the existence for discrete MIM surface plasmon modes. To verify this expectation we plot the analytical dependence $\lambda_\text{plasmon}(\lambda)$ for the MIM mode together with the numerically obtained resonance wavelengths and the plasmon wavelength, determined by the cavity length divided by the respective mode number in Fig.~\ref{fig3}b.

It can be clearly seen that both the odd and even cavity modes follow the analytical and continuous resonance position dependence for an infinite MIM surface plasmon polariton waveguide, proving the MIM-PSPP origin of the observed modes. Slight deviations are due to the phase offset at the cavity ends in terms of a non-zero field elongation similar to the one observed for nanoantenna modes \cite{Novotny2007}. Nevertheless this formalism allows to estimate the required cavity length to observe a desired mode at a predefined wavelength.

%\begin{figure}[htb]
%\centerline{\includegraphics[width=8.0cm]{fig4}}
%\caption{(Color online)  \textbf{(a)} The measured reflection [black solid (TM), triangles (TE)], the statistically simulated (blue dashed) and the spectra of a perfect nanocavity array (red dotted). \textbf{(b)} A scanning electron microscope image of the fabricated sample (processed false-color to visualize the underlying Al wires).}
%\label{fig4}
%\end{figure}

\begin{figure}[p]
\centerline{\includegraphics[width=15cm]{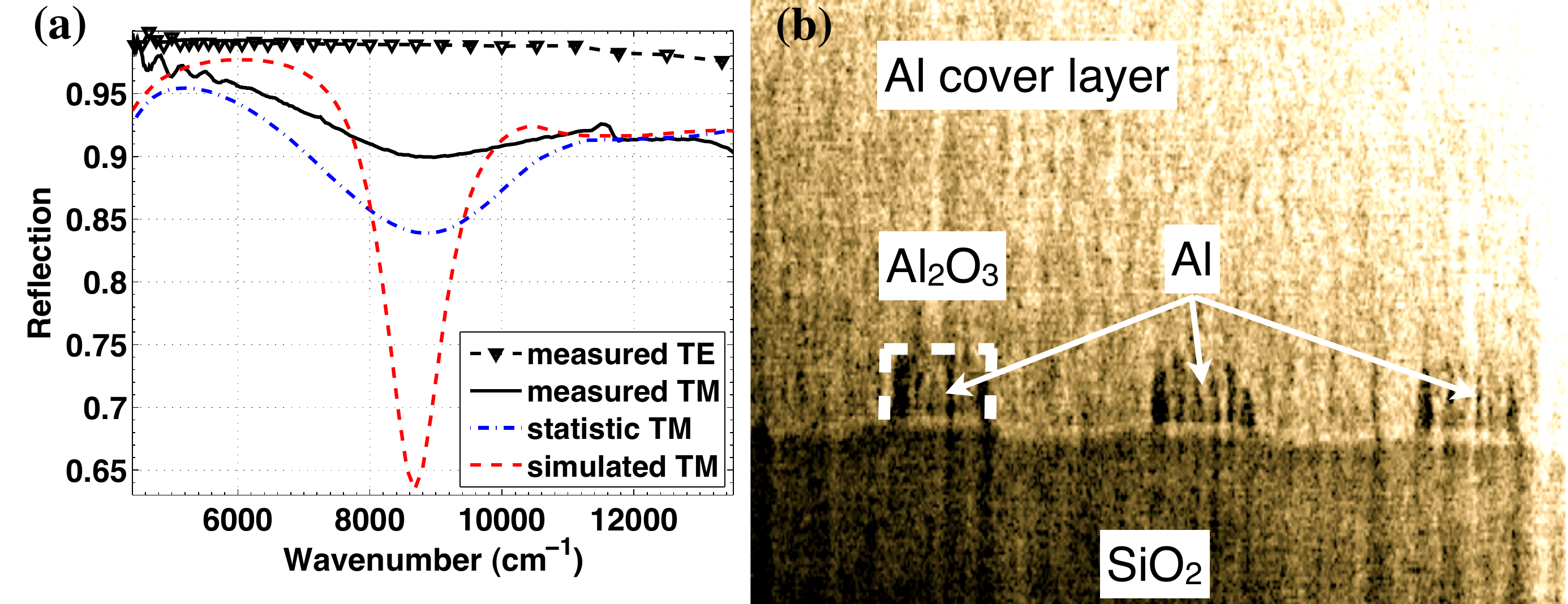}}
\caption{(Color online)  \textbf{(a)} The measured reflection [black solid (TM), triangles (TE)], the statistically simulated (blue dashed) and the spectra of a perfect nanocavity array (red dotted). \textbf{(b)} A scanning electron microscope image of the fabricated sample (processed false-color to visualize the underlying Al wires).}
\label{fig4}
\end{figure}

Finally we want to experimentally prove the spectral observability of such cavity modes. Therefore a demonstrator, schematically shown in Fig. \ref{fig1} has been fabricated for a nanocavity array with a period of $\Lambda_x = 300~\text{nm}$, a height of $h = 60~\text{nm}$, and a width of $b=~170~\text{nm}$. An SEM image of the realized structure is shown in Fig. \ref{fig4}a. The array was optically probed by means of reflection measurements with a commercially available far field spectrometer (\textit{Perkin Elmer} $\lambda950$) in the spectral region of $4000-14000~\text{cm}^{-1}$ ($2.5~-0.7~\mu\text{m}$). The structured sample size was $3\times 3~\text{mm}^2$ yielding a total number of $10^4$ nanocavities within the measured area. The resulting measured reflection shows a dip associated with the $4 th$ mode (Fig. \ref{fig4}b) that appears much weaker compared with theoretical predictions. The reason for this washed-out resonance is the inhomogeneous broadening induced by structural imperfections and surface corrugations \cite{Bernabeu2008}. In contrast to nanostructures with typical dimensions of $10-100~\text{nm}$ the surface fluctuations here are in the same order of magnitude as the $Al_2O_3$ layer thickness yielding a much stronger impact on the spectral observables. Hence, all resonance features appears weaker especially higher order resonances as the one observed here. In order to model this effect the thickness of the oxide layer  was statistically varied  in numerical simulations with a Gaussian distribution around the mean thickness and a standard deviation of $\sigma = 1.4~\text{nm}$. The obtained reflection shows that the resonance position remains fixed but the resonance strength becomes much weaker even for such tiny fluctuations. It can be shown numerically that statistical variations for the entire cavity length cause a similar inhomogeneous broadening but with a weaker influence on the resonance strength (not shown here). Hence, our simulations indicate that surface roughness can be considered as the dominant contribution to spectral broadening effects making a statistical theoretical treatment appropriate for the nanocavity arrays presented here.

In summary we have introduced nanocavitiy arrays based on planar MIM PSPP waveguides. An analytical estimation based on the exact dispersion relation for the MIM configuration permits to predict the resonance positions of the existing modes. The system has been studied for a certain material combination making use of the natural oxidation process of $Al$, but in general the same effects can be observed with other waveguides composed of thicker dielectric gaps and other metals. Experimentally the presence of one particular nanocavity mode could be observed via spectral far field measurements. But due to surface imperfections being in the same order of magnitude as the nanocavity thickness the measured reflection is much weaker as theoretically expected. Especially for a doping of the dielectric gap material with any active material, e.g. fluorescent atoms \cite{Gianini2009} or nonlinear composites \cite{Hui2009}, the cavity setup might be promising for active devices \cite{Oulten2009,Stockman2008} or nonlinear frequency conversion processes. 
Thereby the small mode volume might induce an increased Purcell factor even for the comparably small Q factors observed herein.
\\
Financial support by the Federal Ministry of Education and Research
as well as from the State of Thuringia within the Pro-Excellence
program is acknowledged. %The authors thank for the support of N. Kaiser and his group at the Fraunhofer Institute Jena concerning realistic aluminium-oxide layer thickness measurements.

\newpage

\end{document}